\documentclass[fleqn,usenatbib]{mnras}

\usepackage{newtxtext,newtxmath}

\usepackage[T1]{fontenc}
\usepackage{ae,aecompl}

\usepackage{graphicx}	
\usepackage{amsmath}	
\usepackage{amssymb}	

\title[Cluster Analysis of GRBs in the BATSE Catalogue]{Gaussian-mixture-model-based cluster analysis of gamma-ray bursts in the BATSE catalogue}

\author[B. G. T\'oth et al.]{
B. G. T\'oth,$^{1}$\thanks{E-mail: toth.bence@uni-nke.hu}
I. I. R\'acz$^{1}$ and
I. Horv\'ath$^{1}$
\\
$^{1}$National University of Public Service, Budapest, Hungary\\
}

\date{Accepted XXX. Received YYY; in original form ZZZ}

\pubyear{2019}

\begin{document}
\label{firstpage}
\pagerange{\pageref{firstpage}--\pageref{lastpage}}
\maketitle

\begin{abstract}
Clustering is an important tool to describe gamma-ray bursts (GRBs). We analyzed the Final BATSE Catalog using Gaussian-mixture-models-based clustering methods for six variables (durations, peak flux, total fluence and spectral hardness ratios) that contain information on clustering. Our analysis found that the five kinds of GRBs previously found by other authors are only the cut groups of the previously well-known three types (short, long and intermediate in duration). The two short and intermediate duration groups differ mostly in the peak flux. Therefore, the reanalysis of the BATSE data finds similar group structures than previously. Because the brightness distribution is asymmetric and not correlated with durations or hardnesses the Gaussian mixture model cuts the Short and the Intermediate duration groups into two subgroups, the dim ones and the bright ones.

\end{abstract}




\section{Introduction}

{In recent years statistical clustering tests and pattern recognition algorithms have advanced our understanding of gamma-ray bursts (GRBs), GRB classification, and the relative importance of observed GRB properties.  Statistical clustering tests and pattern recognition algorithms are capable of delineating classes in overlapping parameter spaces and for classification parameters characterized by large measurement uncertainties.

It is difficult to unambiguously classify GRBs even though they are the most luminous explosions in the universe. Each GRB photon contains spectro-temporal information useful to the classification process, and yet detectors can only observe small numbers of GRB photons due to large GRB distances, the transient nature of GRB emission, and inherent instrumental inefficiencies in detecting high-energy photons. As a result, GRB classification parameters have been limited to generic, easily-measured properties that do not do justice to the broad, complex range of individual GRB behaviors. For example, individual GRB light curves exhibit characteristics ranging from smoothly-varying to highly variable and undergo rapid spectral changes, yet existing GRB classification parameters have been almost entirely limited to easily-measured parameters such as duration, spectral hardness, and fluence.

Before the formal application of clustering algorithms to GRB classification, it was recognized that two different GRB classes could explain the GRB duration distribution \citep{maz81,nor84}. The Burst And Transient Source Experiment (BATSE; \citep{m6}) aboard NASA's Compton Gamma-Ray Observatory (CGRO) provided additional justification for this delineation into long and short GRB classes (e.g. \cite{kou93,kos96}). Significant overlap exists in the duration distribution, particularly for GRBs with durations around 2-3$s$. Nonetheless, the two-class structure caught on \citep{nor01,bal03,zha09,luli10,li16} and models involving black hole formation were developed to explain the large luminosities and short emission timescales of all GRBs \citep{woosley17,feng18,fqkc18,SongLiu18}. Standard stellar core collapse models occurred on timescales too long to explain short GRBs, so models involving compact object mergers in binary systems were developed to explain these \citep{paczy86,usov92,peram10,berger14}.

Subsequent evidence supported this simple binary classification system. The host galaxies and redshift distributions of short and long GRBs differ \citep{berger14,levan16}, with the more luminous long GRBs being found in star-forming galaxies. Some low-luminosity long GRBs have been associated with Type Ic supernovae (SN) \citep{hjor03,camp06,pian06,blan16}, supporting the idea that the long GRBs in general are related to deaths of massive stars \citep{woo93,paczy98,wb06,blan16}. For short GRBs, the absence of SN association, the location of these events in metal-poor regions, and their lower luminosities disfavor a massive star origin and point to compact binary mergers \citep{paczy86,usov92,berger14}.

The problem with this binary classification system is that it tempts astronomers to sort all GRBs into either the long or the short GRB class without allowing for the possibility that the classification system is incomplete. Only five years after BATSE observations supported the two-class interpretation, multi- and uni-variate statistical analysis techniques indicated evidence for a third BATSE GRB class \citep{muk98,hor98}. Many authors \citep{hak00,bala01,rm02,hor02,hak03,bor04,hor06,chat07,zito15} have since confirmed the existence of an intermediate GRB class using statistical techniques and/or data mining algorithms. An intermediate class has similarly been identified in Beppo-SAX \citep{hor09} and Swift data \citep{zc07,hor08,huja09,hor10,kb12,tsu14,ht16,tarno16NewA}, even though Beppo-SAX had a smaller effective area and though Swift works in a different energy range than BATSE. The problem of force-assigning each GRB to either the long or short class became apparent with GRB171081A (associated with gravitational wave source GW170817), a burst with intermediate class attributes that physically fits into the merging neutron star scenario \citep{hth18}. In addition to this ambiguous individual case, the two-class scheme fails to recognize that instrumental and sampling biases are capable of creating additional clusters in the observational data, even when physical mechanisms might not be responsible. An example is truncation of the BATSE GRB fluence distribution used in classification caused by the instrument's peak flux trigger that might be responsible for many intermediate GRBs \cite{hak03}. 

In the same way that three GRB classes might provide a more statistically robust description of the data than two classes, inclusion of additional classification attributes, other classification techniques, and additional GRB databases are capable of allowing additional classes to be identified. However, the extracted classes need to be analyzed carefully to determine whether or not they represent real source populations or simply additional sampling biases. 

An example of a classification scheme involving more than three classes can be found in the recent analysis of \cite{chat17}. These authors reanalyzed the BATSE catalog and found five GRB classes while searching a multidimensional parameter space using Gaussian mixture modelling (GMM). The ``five" groups found by the authors are recognizable as subdivisions of the recognized long, intermediate, and short GRB classes: one group represents "bright long GRBs", another "dim long GRBs", a third "bright intermediate GRBs", a fourth "dim intermediate GRBs", and a fifth "short GRBs."  However, no explanation was provided as to what these five classes represent. In this paper we will attempt to reproduce the \cite{chat17} results with the hope of explaining how their five classes are related to the three statistically-accepted GRB classes and the two known GRB source populations.}

The paper is organized as follows. Section 2 discusses the properties of the Fermi GBM catalog, Section 3 describes the classification process, Section 4 discusses the results and Section 5 provides the paper's conclusions.

\section{The current BATSE catalog}

The current BATSE catalog\footnote{https://heasarc.gsfc.nasa.gov/w3browse/all/batsegrb.html} contains 2702 GRBs in total.
	For our analysis, the $T_{50}$, $T_{90}$, $F_t$, $H_{32}$, $H_{321}$ and $P_{256}$ parameters of the GRBs are used.
	$T_{50}$ and $T_{90}$ are the time by which 50\% and 90\% of the total flux of the bursts arrive, respectively. $F_t$ is the total fluence, calculated by summing the time integrated fluences in the 20-50 keV, 50-100 keV, 100-300 keV and >300 keV spectral channels (denoted by $F_1$, $F_2$, $F_3$ and $F_4$): $F_t = F_1 + F_2 + F_3 + F_4$. The $H_{32}$ and $H_{321}$ are the spectral hardness ratios calculated as $H_{32}=F_3/F_2$ and $H_{321}=F_3/(F_2+F_1)$. The $P_{256}$ is the peak flux measured in bins of 256 milliseconds.
	Some bursts have 0 value for some of the fluence values, which means that in these channels the bursts cannot be distinguished from the background as the fitted background variation is in the same order of magnitude as the time integrated fluence. Sometimes this leads to negative fluence values which are then substituted by zero. If only $F_4$ is equal to zero for a burst and so the hardness ratios can be calculated, the total fluence is calculated as $F_t = F_1 + F_2 + F_3$. Therefore, if $F_4$ is zero the $F_t = F_1 + F_2 + F_3 + F_4$ definition still holds.
	In a recent paper by Chattopadhyay and Maitra \citep{chat17}, the bursts with $F_4 = 0$ were excluded from the data analysis and thus only for 1599 bursts were taken into account. In our study, we include the GRBs with $F_4 = 0$, i.e. the full sample of 1929 bursts is used.

For our analysis, we used the R language and environment \citep{rlang}.

\subsection{Outliers}
\label{sec:maths} 

The sample should always be tested for outliers and if any of them is detected, then should be removed. For this task, the \textit{HDoutliers()} function of the R package HDoutliers was used \citep{hdout}. The function fits an exponential distribution to the upper tail of the nearest-neighbor distances between the elements in the sample. If an observation falls in the (1-\textit{alpha}) tail of the fitted cumulative distribution function, it is considered an outlier. The manual of this package suggests to set the threshold parameter \textit{alpha} to 0.05 but increasing the parameter up to 0.25 still no outliers are found.

\subsection{Correlation}

\begin{figure}
	\includegraphics[width=\columnwidth]{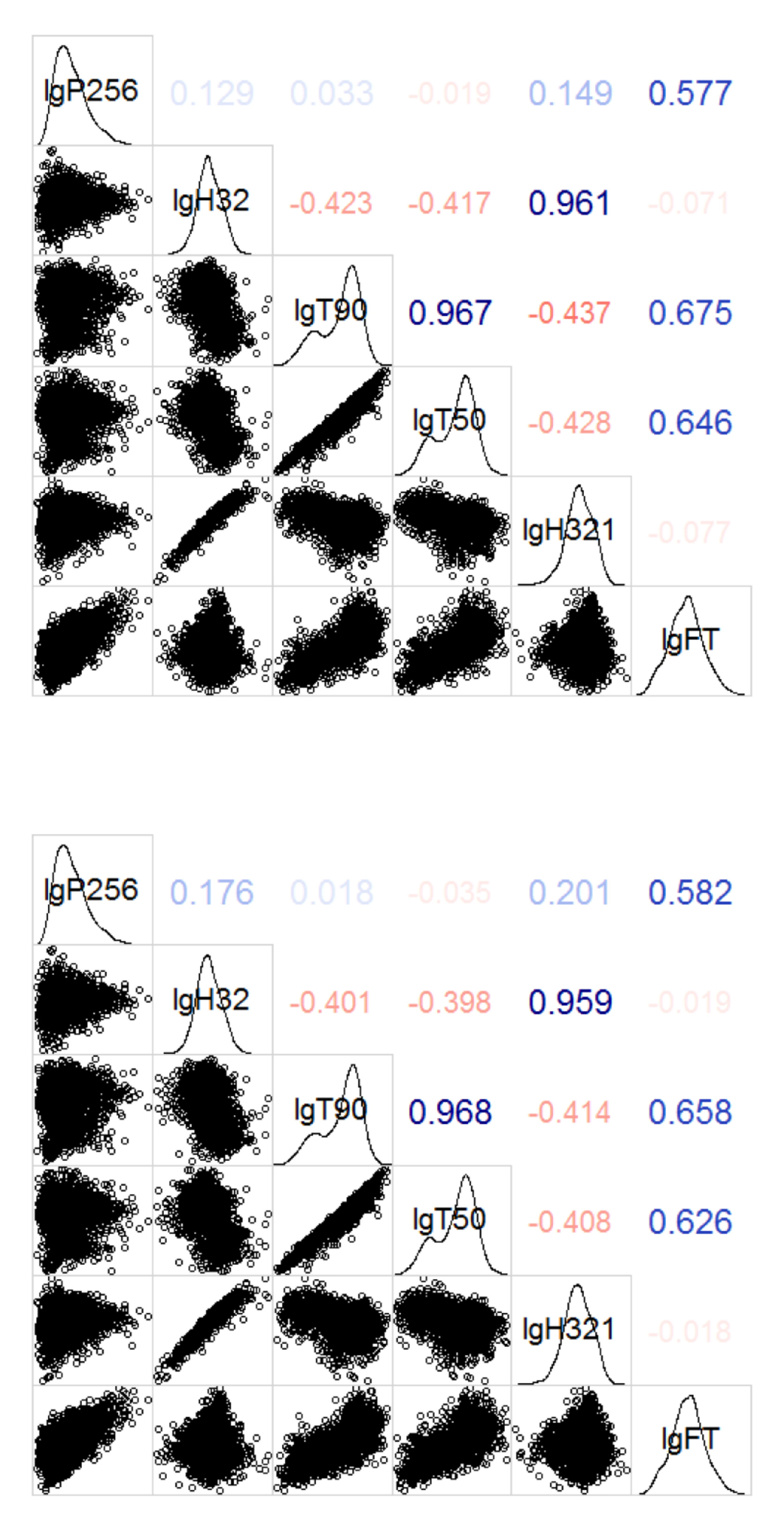}
    \caption{Density functions, 2D plots and Pearson correlation of the variables for the 1599-burst 6-variable case (top) and for the 1929-burst 6-variable case (bottom). The correlogram for the 1929-burst 5-variable case is the top left 5 $\times$ 5 part of the 1929-burst 6-variable case.}
    \label{fig:fig1}
\end{figure}

The correlation coefficients were calculated by the Pearson, Kendall, and Spearman methods of the \textit{corrgram()} function of the R package corrgram \citep{corrgram}. The values obtained by the Kendall method differ in average more than 15\% from the other two, thus the values given by the Pearson method are used. The numerical values for the 1599 GRBs without zero $F_4$ value using all six variables and for all 1929 GBBs with all six variables along with the density functions and the 2D plots of the variables can be seen in Fig.~\ref{fig:fig1} along with the 2D scatterplots for all pairs of variables. The 1929-burst 5-variable case, the one excluding the $F_t$ variable, is actually the top left 5 $\times$ 5 part of the latter. One can see the obvious high correlation between the $T_{90}$ and the $T_{50}$ variables and also the $H_{32}$ and the $H_{321}$ variables in all cases.

\subsection{Seriating the variables}

\begin{figure}
	\includegraphics[width=\columnwidth]{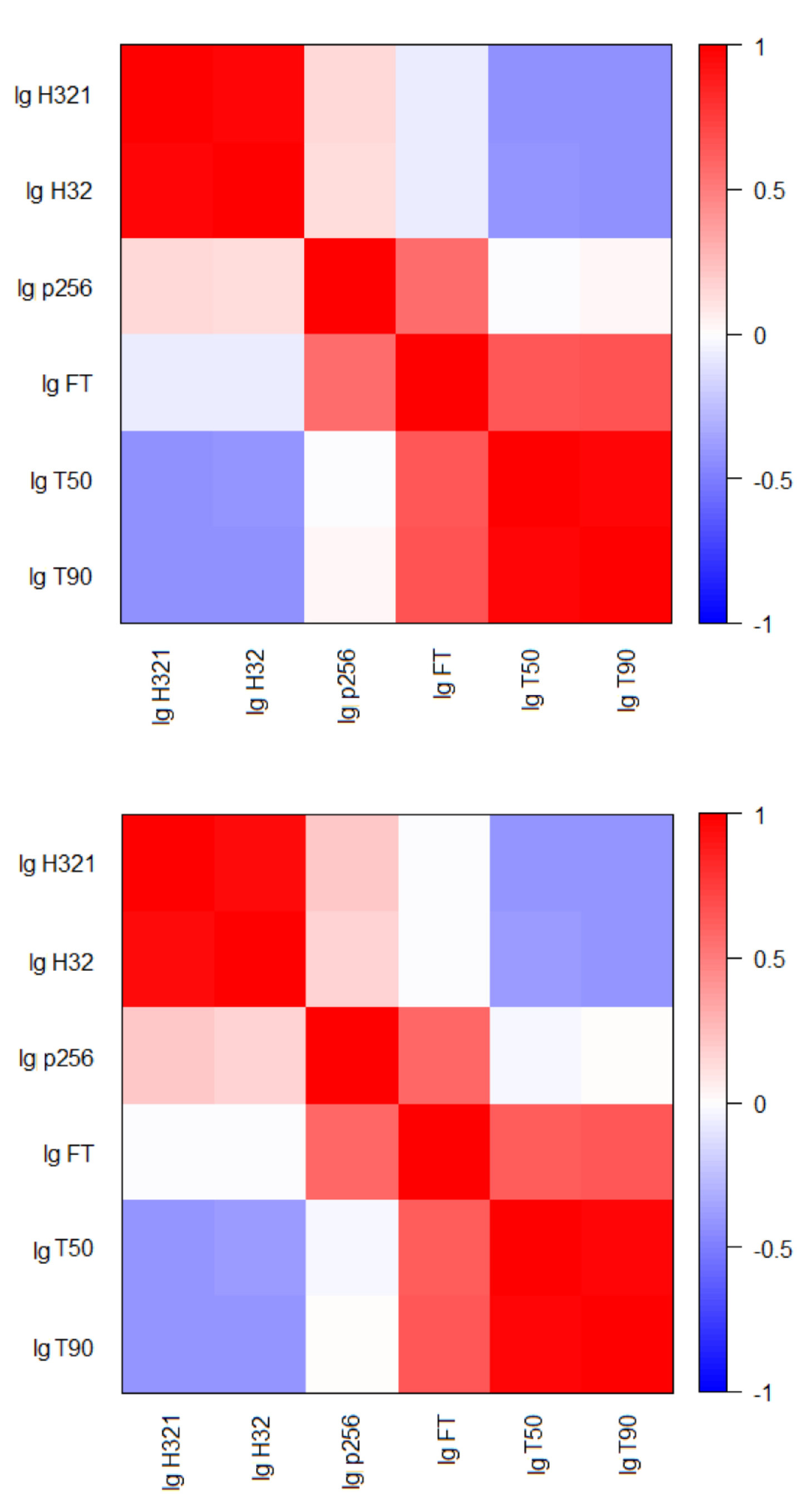}
    \caption{The correlation matrices constructed with the Pearson method and seriated (blockdiagonalized) with the PCA method for the 1599-burst 6-variable case (top) and for the 1929-burst 6-variable case (bottom). The 1929-burst 5-variable case is part of the 1929-burst 6-variable case.}
    \label{fig:fig2}
\end{figure}

The six chosen variables, however, have some information overlap on the attributes of the GRBs themselves. Furthermore, the three attributes that contain the gratest amount of orthogonal information are the duration, the hardness and the intensity. To obtain the measures of these attributes as clearly as possible, seriation was performed on the whole dataset with the R package seriation \citep{seriation} using the function \textit{seriate()} and the methods PCA and PCA\_angle on the correlation matrices calculated by using the Pearson, the Kendall and the Spearman methods. We use these three methods, because all of them have different advantages. The PCA method projects the data on its first principal component to determine the order, while the PCA\_angle method projects the data on the first two principal components and then orders by the
angle in this space. The results obtained from the two techniques are very similar to one another, though the actual values of the correlation matrix calculated by using the Kendall method differs from the other two. The color map plot of the seriated correlation matrix constructed with the Pearson method and seriated with the PCA method is shown in Fig.~\ref{fig:fig2}, again for the 1599-burst 6-variable case and the 1929-burst 6-variable case, the 1929-burst 5-variable case being part of the latter.

	In all cases, the matrices show the same behavior: the $T_{50}$ and $T_{90}$ values are strongly correlated with each other, as well as the $H_{32}$ and the $H_{321}$ values and these two 2-element blocks are anti-correlated with each other. The $F_{t}$ variable shows correlation with the time variables, and the $P_{256}$ shows correlation with the fluence variables, but neither correlates with the other two element block. However, the $F_{t}$ and the $P_{256}$ variables are also correlated with each other, forming an overlapping block between the two 2-element blocks.

\section{Number of clusters}

The R package Mclust \citep{mclust} was used to determine the value of the Bayesian Information Criterion (BIC) using the function \textit{Mclust()}. There are 14 different built-in models to fit ellipsoidals on the data points in the space with dimensions equal to the number of the used variables. The models use the combination of four parameters. The first is if the fitted distribution is spherical or ellipsoidal. The second and the third applies only if for ellipsoidals: the second is if the axes are parallel to the coordinate axes, aligned with each other or independent of each other and the third is if the shape of all fitted ellipsoidals are the same or not. The fourth parameter is if the fitted spherical or ellipsoidal distributions all have the same volume or not. For detailed description see \cite{mclust}.

The BIC values were calculated using all 14 methods and assuming components between 1 and 10. In the 1599-burst 6-variable case, the negative BIC values provided by the EII, VII, EEI, VEI, EVI, VVI and VVE models were more than 800 larger than the ones obtained with the other methods.
Figure 5 of \citet{chat17} contains only one BIC function and, if we study carefully our curves, the one obtained by the EVV model (ellipsoidals of equal volume) must be the one in their figure. However, the extremal BIC value of -2357 is reached at five assumed groups with the EVV model, which is smaller than the extremal BIC value obtained by the VVV model (fitting ellipsoidals with varying volume, shape, and orientation) by assuming 4 groups, which is -2295. Even the BIC value provided by the VVV model with four assumed groups is higher than the extremal value of the ones by EVV model. This leads to the result, that the VVV model has to be used in this case to determine the optimal number of clusters.

\begin{figure}
	\includegraphics[width=\columnwidth]{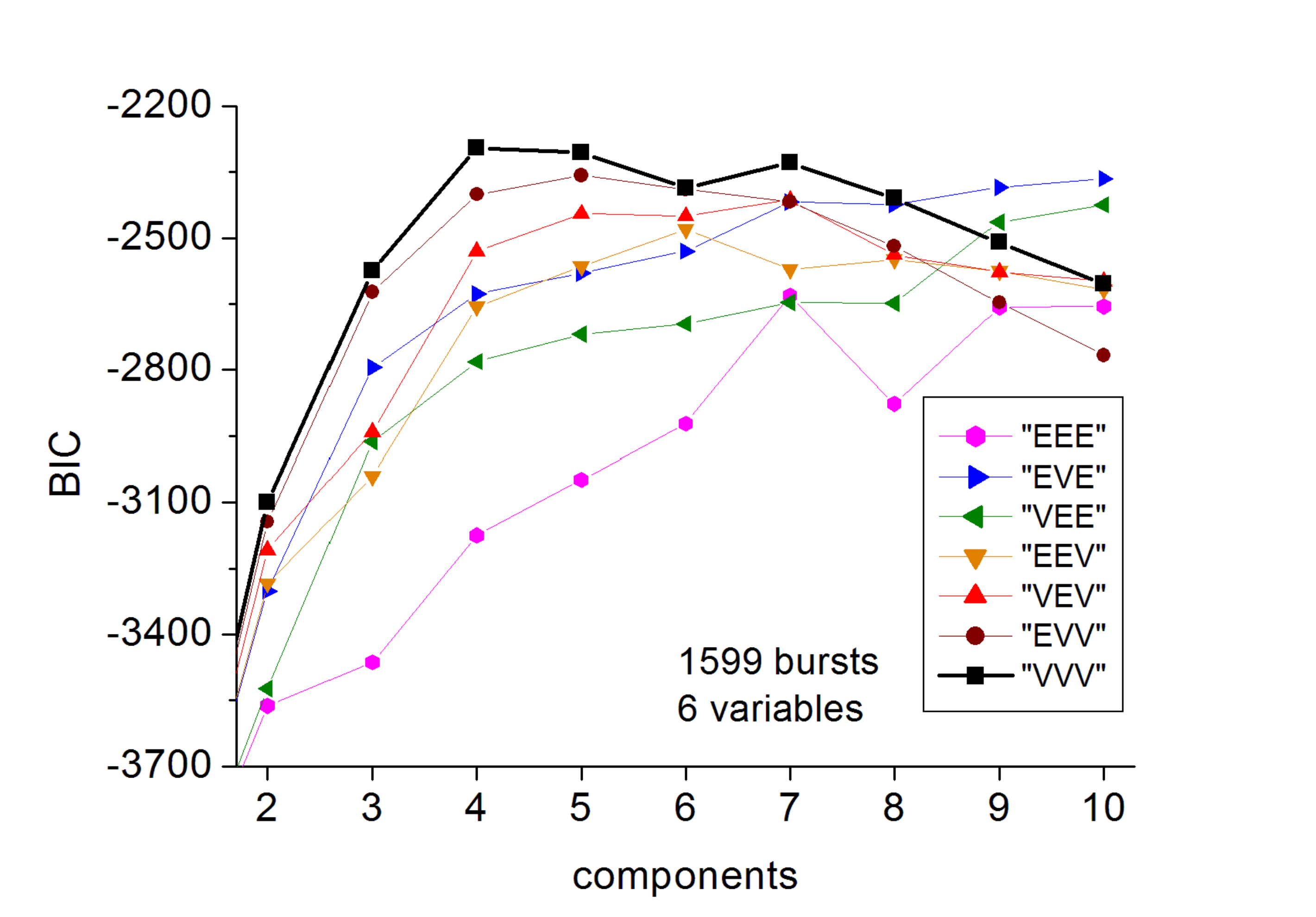}
    \caption{The Bayesian Information Criterion (BIC) for 1599 GRBs (without the ones with no F4) and 6 variables (lg $T_{50}$, lg $T_{90}$, lg $H_{32}$, lg $H_{321}$, lg $P_{256}$ and lg $F_t$) for the models showing a maximum assuming 2 to 10 GRB groups.}
    \label{fig:fig3}
\end{figure}

\begin{figure}
	\includegraphics[width=\columnwidth]{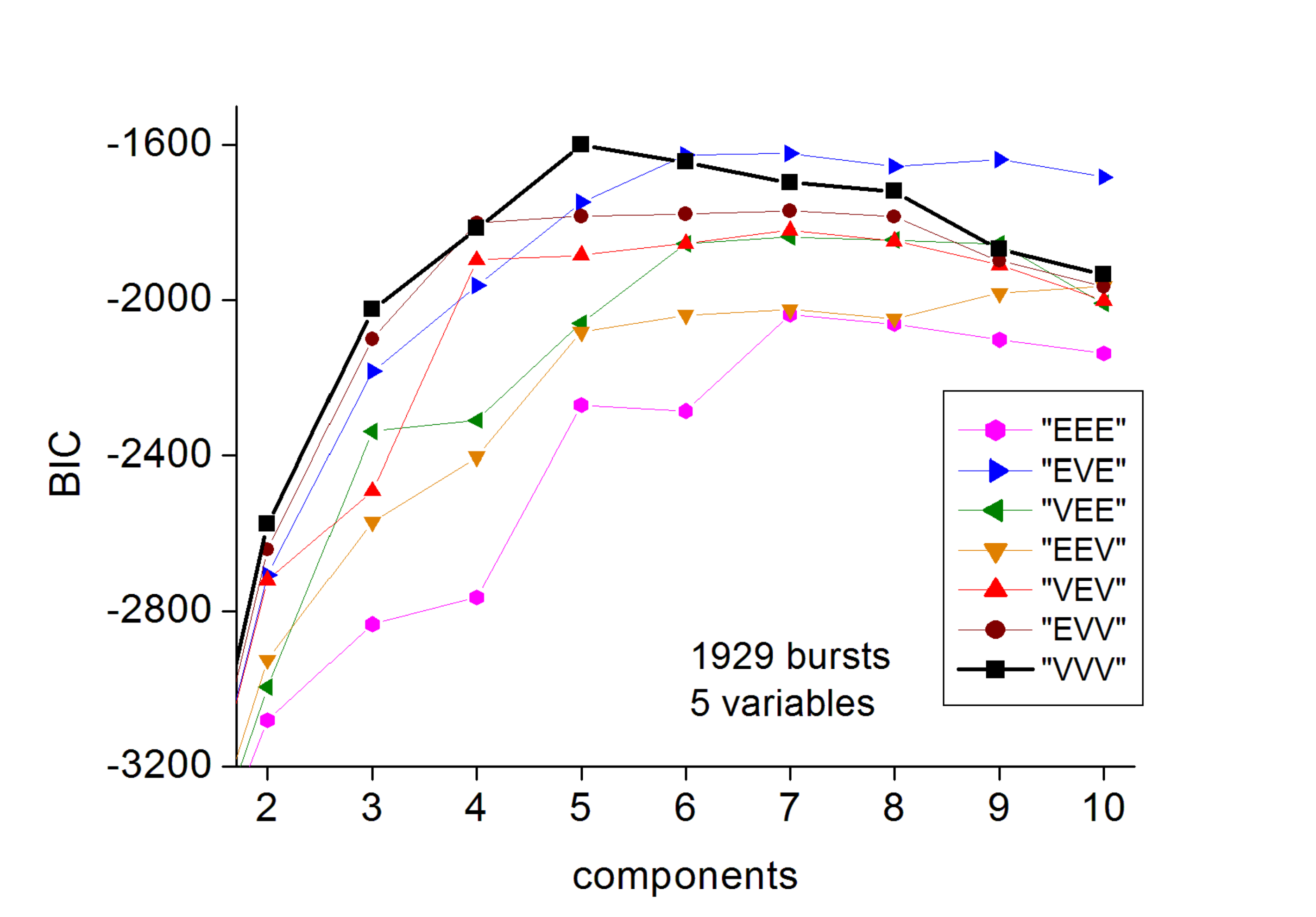}
    \caption{The Bayesian Information Criterion (BIC) for all 1929 GRBs and 5 variables (lg $T_{50}$, lg $T_{90}$, lg $H_{32}$, lg $H_{321}$ and lg $P_{256}$) for the models showing a maximum assuming 2 to 10 GRB groups.}
    \label{fig:fig4}
\end{figure}

	In the 1929-burst 5-variable case, the VVV method provides the extremal BIC value for 5 assumed groups among all 14 models (Fig.~\ref{fig:fig4}, the seven BIC functions obtained by the other models with much more negative values are omitted). The mean value of each variable for the five groups are shown in Table~\ref{tab:table1}. It can be clearly seen from the $T_{90}$ and $T_{50}$ values that the Intermediate (groups \#1 and \#3) and the Long (groups \#2 and \#4) have been split into two: a dimmer (groups \#1 and \#2) and a brighter (groups \#3 and \#4) one. The hard Short GRBs remain together.
	
\begin{table}
	\centering
	\caption{The mean values of the 5 variables in each of the 5 groups in the 1929-burst case provided by the \textit{mclust()} function. $N$ is the population of the group.}
	\label{tab:table1}
	\begin{tabular}{ c | c | c | c | c | c | c | c }
		  & lg $P_{256}$ & lg $H_{32}$ & lg $T_{90}$ & lg $T_{50}$ & lg $H_{321}$ & $N$ \\ \hline
        1 & 0.675 & 0.457 & 0.669 & -0.055 & 0.219 & 182 \\ \hline
        2 & 0.457 & 0.464 & 1.610 & 1.119 & 0.221 & 511 \\ \hline
        3 & -0.120 & 0.560 & 0.322 & -0.110 & 0.189 & 130 \\ \hline
        4 & -0.059 & 0.366 & 1.470 & 1.093 & 0.096 & 761 \\ \hline
        5 & 0.198 & 0.769 & -0.298 & -0.639 & 0.592 & 345 \\
	\end{tabular}
\end{table}

\begin{figure}
	\includegraphics[width=\columnwidth]{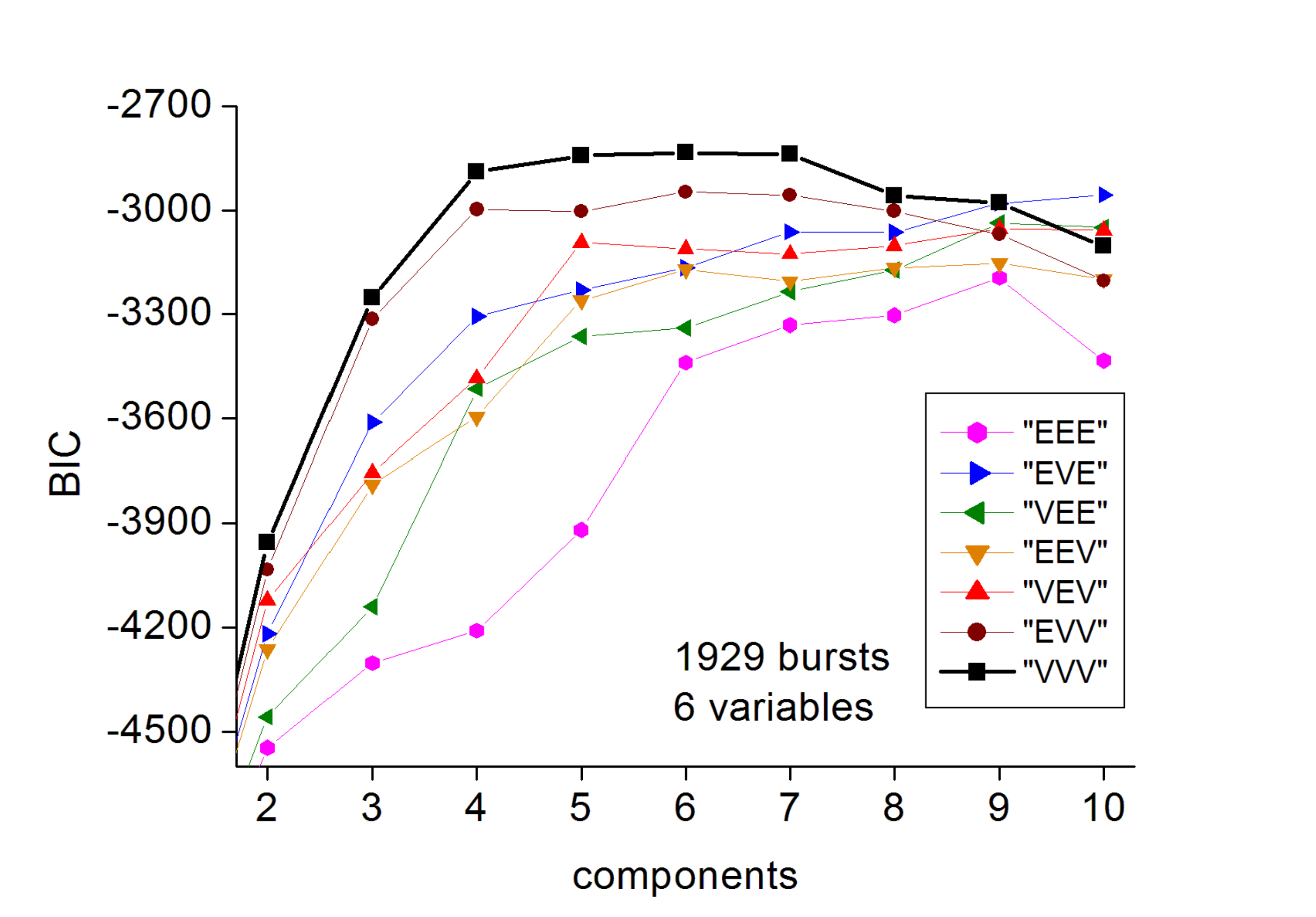}
    \caption{The Bayesian Information Criterion (BIC) for all 1929 GRBs and 6 variables (lg $T_{50}$, lg $T_{90}$, lg $H_{32}$, lg $H_{321}$, lg $P_{256}$ and lg $F_t$) for the models showing a maximum assuming 2 to 10 GRB groups.}
    \label{fig:fig5}
\end{figure}

	The highest BIC value for the 1929-burst 6-variable case, which we regard as being the correct base of this analysis, was also obtained by the VVV method (Fig.~\ref{fig:fig5}). However, there is no clear maximum, the function has a plateau between 4 and 7 groups, but all these values lie clearly above the next highest values, obtained by the EVV method.
 
    If we accept the result of 5 groups of the previous two cases, the mean value of the six variables for the five groups can be calculated using the \textit{mclust()} function. The values can be seen in Table~\ref{tab:table2}.

\begin{table}
	\centering
	\caption{The  mean values of the 6 variables for the 5 groups found to be optimal for the 1929-burst case by using the Bayesian Information Criterion provided by the \textit{mclust()} function. $N$ is the population of the group.}
	\label{tab:table2}
	\begin{tabular}{ c | c | c | c | c | c | c | c }
		  & lg $P_{256}$ & lg $H_{32}$ & lg $T_{90}$ & lg $T_{50}$ & lg $H_{321}$ & lg $F_{t}$ & $N$ \\ \hline
		1 & 0.273 & 0.768 & -0.266 & -0.617 & 0.590 & -6.190 & 312 \\ \hline
		2 & 0.167 & 0.417 & 1.591 & 1.168 & 0.162 & -5.228 & 747 \\ \hline
		3 & 0.728 & 0.438 & 1.275 & 0.651 & 0.188 & -5.045 & 316 \\ \hline
		4 & -0.014 & 0.760 & -0.395 & -0.930 & 0.471 & -6.579 & 148 \\ \hline
		5 & -0.151 & 0.351 & 1.313 & 0.953 & 0.065 & -5.835 & 406 \\
	\end{tabular}
\end{table}

\section{Discussion}

To analyze the data of GRBs, usually the duration, the hardness and the fluence variables are taken into account. The three groups found by using these variables are the hard Short, the softer Long and the Intermediate group. If the peak flux is also incorporated into the analysis, the splitting of the Short and the Intermediate groups can be observed. In Table 2, the lg $T_{90}$ and lg $T_{50}$ values for groups \#1 and \#4 indicate that these GRBs are Short ones, groups \#3 and \#5 are Intermediate ones and \#2 is the Long group.

This is also supported by the values of $H_{321}$: the two Short groups are the hard ones while the other three are much softer. But the two Short GRB groups and also the two  Intermediate groups differ in their $P_{256}$ values significantly: the Short group \#1 and the Intermediate group \#3 are much more luminous than the Short group \#4 and Intermediate group \#5.

However, this does not underpin the presence of subgroups of different physical properties, because the new groups appear due to two reasons. On the one hand, the distribution of the brightness is asymmetric and, on the other, it is uncorrelated to the duration and the hardness variables as was shown previously in Fig.~\ref{fig:fig2}. This leads to the Gaussian mixture model to cut the Short and Intermediate groups into a dim and a bright group.

\section{Conclusions}

In this paper, with multivariate analysis, we analyzed the CGRO BATSE final catalogue data using six variables: two durations ($T_{50}$, $T_{90}$), the total fluence ($F_{t}$), the 256 m$s$ peak flux ($P_{256}$) and two hardness ratios ($H_{32}$, $H_{321}$). Many papers analyzed BATSE data using uni- \citep{hor98,hor02,zito15} or multivariate \citep{muk98,hak00,bala01,rm02,hak03,hor06,chat07,chat17,acuneryde18} analysis concluding that there is also an Intermediate duration class of GRBs along with the common two Short and Long type GRBs. \citet{muk98} using the same six variables ($T_{50}$, $T_{90}$, $F_{t}$, $H_{32}$, $H_{321}$ and $P_{256}$) with two completely independent mathematical procedures (non-parametric hierarchical cluster analysis, model-based maximum likelihood clustering analysis) found a very similar three component structure. \citet{chat07} using $k$-means partitioning method and the Dirichlet process of mixture modelling with also the same six parameters ($T_{50}$, $T_{90}$, $F_t$, $H_{32}$, $H_{321}$ and $P_{256}$) found also three kind of GRBs. \citet{chat17} carried out multi-dimensional analysis of the BATSE data with six parameters but only for 1599 GRBs because of the assumption that if $F_4$ is missing the $F_t$ variable cannot be used (in a recent paper they published similar results \citep{chat18}).
Because of the same reason, all 1929 bursts was analyzed only in a five-parameter space. However, this is not correct since if $F_4$ is not high enough to observe it (to be distinguishable from zero) the definition of the total fluence still has a physical meaning.

In this work, we not only repeated the analyis for the 6-variable 1599-burst and the 5-variable 1929-burst case, but also carried out for the 6-variable 1929-burst case. Our results are similar to \citet{chat17} even when the extremal BIC value of the fitting resulted from another model: the fitting of ellipsoidals with varying volume, shape and orientation.

From these results, the conclusions can be improved as one can compare the results of this paper with previous results. The most commonly used categorization of GRBs is based on their duration: as previously the Short and Long GBRs had been recognized the third group was named Intermediate group. In this paper our analysis found two Short, two Intermediate and one Long group the two Short and Intermediate duration groups differing mostly in the peak flux.
Because the brightness ($P_{256}$) distribution is asymmetric and not correlated with the duration or hardness variables (see Fig.~\ref{fig:fig1}) the Gaussian mixture model cuts the Short and Intermediate duration groups into two parts, a dim and a bright one. By fitting them with a symmetric function (the Gaussian distribution) one can get a better fit by cutting the asymmetric distribution into two parts. 
This means that there are no five subgroups but the splitting of the Short and Intermediate groups into two parts is an effect of the $P_{256}$ variable on the clustering. This effect of the brightness variable on the classification has to be taken into consideration in further analyses.

\section*{Acknowledgements}

The authors are grateful to Z. Bagoly and L.G. Bal\'azs for the useful advice. The authors thank the Hungarian TIP and TKP program for their support.  
The authors gratefully thank for J. Hakkila  for useful comments
and recommendations which improved the paper.







\hyphenation{Post-Script Sprin-ger}





\bsp	
\label{lastpage}
\end{document}